\documentclass{revtex4}

\usepackage[utf8]{inputenc} 
\usepackage{amsmath}
\usepackage{amssymb}
\usepackage{mathrsfs}
\usepackage{bm}
\usepackage{color}
\usepackage{braket}
\usepackage{subfig}
\usepackage{graphicx}

\DeclareGraphicsExtensions{.pdf,.jpg}

\newcommand{\tr}{\mbox{Tr}}
\renewcommand{\bf}[1]{\textbf{#1}}

\begin{document}
\author{Syed Tahir Amin$^{1,2,4}$, Bruno Mera$^{1,3}$, Nikola Paunkovi\'c$^{1,3}$, and V\'{\i}tor R. Vieira$^{2,4}$}

\affiliation{ $^1$ Instituto de Telecomunica\c{c}\~oes, Av. Rovisco Pais 1, 1049-001 Lisbon, Portugal}
\affiliation{$^2$  Departamento de F\'{\i}sica, Instituto Superior
T\'ecnico, Universidade de Lisboa, Av. Rovisco Pais 1, 1049-001 Lisboa, Portugal}
\affiliation{$^3$ Departamento de Matemática, Instituto Superior Técnico, Universidade de Lisboa, Av. Rovisco Pais 1, 1049-001 Lisboa, Portugal}
\affiliation{$^4$ CeFEMA, Instituto Superior T\'ecnico, Universidade de Lisboa, Av. Rovisco Pais 1, 1049-001 Lisboa, Portugal}

\title{Information geometric analysis of long range topological superconductors}

\begin{abstract}
The Uhlmann connection is a mixed state generalisation of the Berry connection. The latter has a very important role in the study of topological phases at zero temperature. Closely related, the quantum fidelity is an information theoretical quantity which is a measure of distinguishability of quantum states. Moreover, it has been extensively used in the analysis of quantum phase transitions. In this work, we study topological phase transitions in $1$D and $2$D topological superconductors with long-range hopping and pairing amplitudes, using the fidelity and the quantity $\Delta$ closely related to the Uhlmann connection. The drop in the fidelity and the departure of $\Delta$ from zero signal the topological phase transitions in the models considered. The analysis of the ground state fidelity susceptibility and its associated critical exponents are also applied to the study of the aforementioned topological phase transitions.  
\end{abstract}
\maketitle

\section{Introduction}

The last decade saw a remarkable development in the research in topological quantum matter because of its unique properties which are in general immune to small local perturbations. Topological phases (TPs) of matter, like topological insulators and superconductors, are described in the bulk by topological invariants, such as the winding number of a map or the Chern number of a vector bundle over the Brillouin zone~\cite{tho:koh:nig:den:82}, which justify the robustness to perturbations preserving the gap. By changing the parameters of the Hamiltonian, such as hopping amplitudes or external fields, we can control the phase of the system by closing the gap. When this gap closing is accompanied by a change in the values of topological invariants we say that we have a topological phase transition. Such phase transitions are witnessed by using tools from quantum information such as entanglement~\cite{ost:ami:fal:faz:02,su:son:gu:06,vid:lat:ric:kit:03,oli:sac:14} and, more recently, the ground state fidelity and the associated information geometric notions~\cite{uhl:76,alb:83,zan:qua:wan:sun:06,zan:pau:06,pau:sac:nog:vie:dug:08,sac:pau:vie:11}. The bulk-to-boundary principle states that, if we have a system composed of two parts, described in their bulks by different values of a topological invariant, there will exist gapless excitations localised at their boundary. These so-called edge states are topologically protected.

In~\cite{kit:01}, Kitaev introduced a 1D model of a topological superconductor, with hopping and pairing amplitudes between nearest neighbours; where topological phase transitions were mainly controlled by the chemical potential $\mu$. Since then, a number of modifications with long-range hopping and pairing amplitudes have been considered~\cite{pat:neu:pac:17,pie:gla:von:13,kli:sta:yaz:los:13,pie:gla:leo:von:14,neu:yaz:ber:16,niu:chu:hsu:man:rag:cha:12,vod:lep:erc:gor:pup:14,dut:dut:17,ale:del:17,lep:del:17,viy:vod:pup:del:16}. It was suggested that a long-range superconducting Hamiltonian may be achieved by using a quantum simulator such as an optical lattice with cold atoms, where changing the depth of lattice potential can mimic the exponential decaying hopping amplitudes~\cite{viy:lia:del:ang:18}. There  may be the possibility of realising such Hamiltonians in the Shiba chain~\citep{rub:hei:pen:von:fra:17,deu:car:joe:mig:nic:jos:17}. In the case of Kitaev's model the topological non-trivial phase occurs for $\mu\in(-1,1)$. However, as the long-range amplitudes are switched on, the topological non-trivial phase starts to expand in phase diagram, as shown in Fig.~(1) of~\cite{viy:vod:pup:del:16}. On the other hand, in the case of 2D p-wave superconductors with long-range amplitudes, the topological phases start to either expand or shrink, as shown in Fig.~(1) of \cite{viy:lia:del:ang:18}. The fidelity and the related quantities, such as the ground state fidelity susceptibility and the Uhlmann quantity, are shown to be sensitive and to detect topological phase transitions~\cite{mer:vla:pau:vie:17,mer:vla:pau:vie:17:qw,ami:mer:val:pau:vie:18} and the traditional symmetry breaking phase transitions~\cite{pau:vie:08}. Thus, we want to extend the use of above mentioned quantities to probe topological phase transitions in the long-range models given in \cite{viy:vod:pup:del:16}, for 1D, and \cite{viy:lia:del:ang:18} for the 2D case.

Our paper proceeds as follows. In Section II, we briefly present the expressions for fidelity, the Uhlmann quantity $\Delta$ and the ground state fidelity susceptibility, and explain how these quantities are used to probe topological phase transitions. In Section III, we present the results of the fidelity and Uhlmann connection analysis along with the ground state fidelity susceptibility for $1$D and $2$D TSCs with long-range amplitudes. We also analyse the critical behaviour at zero temperature by looking at critical exponents. Finally, in Section IV, we present our conclusions.

\section{Information geometry and phase transitions}

In this section, we present the main quantities through which topological phase transitions are studied. The fidelity quantifies the distinguishability between two mixed states $\rho$ and $\sigma$ and is given by 
\begin{align}
\label{eq:fid}
F(\rho,\sigma) &= \tr\sqrt{\sqrt{\rho}\sigma \sqrt{\rho}}.
\end{align} 
Consider a smooth family of Hamiltonians $\{\mathcal{H}(q)\}_{q\in \mathcal{Q}}$ where $\mathcal{Q}$ is a configuration space of parameters, assumed to be a smooth manifold. For a given temperature $T$, we can consider the associated family of thermal states, in units where $k_{\text{B}}\equiv 1$,
\begin{align}
\rho(q)=\frac{\exp\big(-\frac{\mathcal{H}(q)}{T}\big)}{Z(q)},\ q\in \mathcal{Q},
\end{align}
where $Z(q)=\tr\big[\exp\big(-\mathcal{H}(q)/T\big)\big]$. One can then consider the fidelity between two close points $q$ and $q'=q+\delta q$. When the two points are within the same phase, the corresponding states $\rho(q)$ and $\rho(q+\delta q)$ will be almost the same and the fidelity will be close to $1$. However, whenever there is a phase transition, the state of the system changes dramatically and the fidelity will drop, signalling the phase transition. In the \emph{thermodynamic limit}, the fidelity between states from two different phases becomes zero. Moreover, generically, in the thermodynamic limit any two distinct states become orthogonal, even when taken from the same phase (consequence of the famous Anderson orthogonality catastrophe~\cite{and:67}). This is why we consider the finite size case, in which the fidelity is not zero since it is a product of a finite number of generically non-zero terms. Note though that this is not necessarily so. Indeed, it is known that for different phases of topological band insulators the corresponding ground state fidelity vanishes exactly~\cite{gu:sun:16} (see~\cite{hua:bal:16} for further study). In the thermodynamic limit, the quantity to be considered is the \emph{fidelity susceptibility}, which is finite in the gapped phases, see Eq.~\eqref{eq:zero_temp_fidsusc} below, for the systems considered.

In our study, we are interested in zero temperature quantum phase transitions. Thus, one might analyse the pure state fidelity in terms of the ground states $\ket{\psi(q)}$, given by 
\begin{align}
\label{eq:fid_pure}
F(\ket{\psi(q)},\ket{\psi(q+ \delta q)}) &= |\langle \psi (q)|\psi (q+\delta q)\rangle|.
\end{align}
For the systems considered, the ground state manifold is non-degenerate outside the set of gapless critical points. One could thus use the above pure state expression to evaluate the ground state fidelity. Nevertheless, since in our case even for finite system size the ground state becomes degenerate at the gapless points, the pure state fidelity can be probed only {\em in the vicinity} of the critical points. To evaluate the fidelity at the critical points as well, we thus consider the zero temperature limit of the  above mixed state expression~\eqref{eq:fid}. Note that at the critical points, the zero temperature limit of the thermal state is not pure, but the (normalised) projector onto the {\em degenerate} ground state manifold.

The notion of state distinguishability quantified by the fidelity provides associated information geometric quantities, namely, a distance and a Riemannian metric in the space of states and a connection in the fibre bundle of purifications~\cite{chr:jam:12}. Consider, for simplicity, a finite dimensional Hilbert space $H$ which we can identify with $\mathbb{C}^n$. For a fixed rank $k$, the space of density operators is a manifold. The so-called Bures distance in this space is defined by $d^2_{B}(\rho,\sigma)=\big(1-F(\rho,\sigma)\big)$. Its infinitesimal version, $d^2_{B}(\rho,\rho+\delta\rho)$ to second order in $\delta\rho$, defines a Riemannian metric. By taking the pullback with respect to the map $q\mapsto \rho(q)$, we obtain the fidelity susceptibility $\chi_{F}(q)$, which is a positive semidefinite rank-$2$ covariant tensor in $\mathcal{Q}$. In this work, we are considering TSCs at $T=0$, in dimensions $D=1,2$, whose Hamiltonian can be expressed as
\begin{align}
\label{eq:HBDG}
	\mathcal{H}(q)= \frac{1}{2}\int_{\text{B.Z.}} \frac{d^{D}k}{(2\pi)^D} \widetilde{\psi}^{\dagger}(\bf{k}) H_{\text{BdG}}(\bf{k};q)\widetilde{\psi}(\bf{k})=\frac{1}{2}\int_{\text{B.Z.}} \frac{d^{D}k}{(2\pi)^D} \widetilde{\psi}^{\dagger}(\bf{k}) d^{\mu}(\bf{k};q)\sigma_{\mu} \widetilde{\psi}(\bf{k}),
\end{align}
where $\widetilde{\psi}^{\dagger}(\bf{k})=(c^{\dagger}(\bf{k})\ c(-\bf{k}))$, in which $c^{\dagger}(\bf{k})$ ($c(\bf{k})$) creates (destroys) an electron with momentum $\bf{k}\in \text{B.Z.}$, where $\text{B.Z.}$ denotes the first Brillouin zone, $\{\sigma_{\mu}\}_{\mu=1}^{3}$ are the standard Pauli matrices and we employ the Einstein summation convention. The topological phase, for fixed $q\in\mathcal{Q}$, is specified by the homotopy class of a map $\Phi_{q}:\text{B.Z.}\to S^D$ given by
\begin{align}
	\Phi_q: \bf{k}\mapsto n^{\mu}(\bf{k};q)=\frac{d^{\mu}(\bf{k};q)}{|d(\bf{k};q)|},
\end{align}
where in the case $D=1$, due to symmetry considerations, the target space is the equator of the Bloch sphere, therefore, a circle $S^1$. The norm of the vector $d^{\mu}(\bf{k};q)$ is the gap function, $|d(\bf{k};q)|\equiv E(\bf{k};q)$. When the gap closes, the map $\Phi_{q}$ stops being well-defined. The expression of the ground state fidelity susceptibility is given by~\cite{ami:mer:val:pau:vie:18}
\begin{eqnarray}
\chi_{F}=\chi_{\alpha\beta}(q)dq^{\alpha}dq^{\beta}=\frac{1}{4}\int_{\text{B.Z.}} \frac{d^Dk}{(2\pi)^D} \delta_{\mu\nu}\frac{\partial n^{\mu}}{\partial q^{\alpha}}\frac{\partial n^{\nu}}{\partial q^{\beta}} dq^{\alpha} dq^{\beta}.
\label{eq:zero_temp_fidsusc}
\end{eqnarray}
Note that the above thermodynamic limit expression for the fidelity susceptibility is not the zero temperature limit of the pullback of the Bures metric. Instead, it is the pullback of the Fubini-Study metric, i.e., the natural Riemannian metric in the space of pure states. This is done because the thermodynamic limit of the former, for quadratic Hamiltonians, does not see the phase transitions~\cite{ami:mer:val:pau:vie:18}.

In addition, one can consider purifications of density matrices. A purification of a density matrix $\rho$ is as a $n\times k$ matrix $w$, such that $\rho=ww^{\dagger}$. There is a $\mbox{U}(k)$ gauge degree of freedom in choosing $w$, since $w U$, with $U\in \mbox{U}(k)$, defines the same density matrix $\rho$. This gauge ambiguity defines the aforementioned (principal) fibre bundle of purifications over the space of density matrices of rank $k$. The fidelity induces a connection in this principal bundle known as the Uhlmann connection~\cite{uhl:89}.

To probe the Uhlman connection, we study the quantity $\Delta $ defined by
\begin{align}
\Delta(\rho(q),\rho(q+\delta q)) :=  
F(\rho(q),\rho(q+\delta q))- \tr(\sqrt{\rho(q)}\sqrt{\rho(q+\delta q)}).
\label{eq:deltafid}
\end{align} 
One can show that,
\begin{equation}
\Delta(\rho(q),\rho(q+\delta q)) = \tr\{|\sqrt{\rho(q+\delta q)}\sqrt{\rho(q)}|(I-V)\}.
\label{eq:deltatrace}
\end{equation}
where $V$ approximates the infinitesimal parallel transport according to the Uhlmann connection in the gauge specified by $w(q)=\sqrt{\rho(q)}$. Note that, when $V=I$ then $\Delta$ vanishes and, thus, this quantity is probing the non-triviality of the Uhlmann connection. Moreover, whenever the states $\rho(q)$ and $\rho(q+\delta q)$ commute we have $\Delta =0$. In other words, the non-triviality of $V$, and, thus, of the Uhlmann connection, quantifies the change in the eigenbasis of $\rho(q)$. The quantity $\Delta$ quantifies part of the change of a density operator and, therefore, much like fidelity, it probes phase transitions (for more details see \cite{pau:vie:08,mer:vla:pau:vie:17,ami:mer:val:pau:vie:18}). Note though that the fidelity captures the total change in the density operator, i.e., the change in {\em both} the eigenvalues and the eigenvectors of the density operator. Therefore, the fidelity is sensitive to a broader range of variations. Nevertheless, $\Delta$ provides finer information concerning state distinguishability. The analysis of both quantities thus allows us to probe what kind of variations of the density operator lead to phase transitions.

The ground state of the topological phase is obtained by filling the occupied bands below the gap and it is non-degenerate. For the systems considered here, there is a single occupied band and the pullback of the Uhlmann connection by the family of single particle states parametrised by momenta is Abelian. The resulting connection is precisely the so-called Berry connection.

We end this section with a remark on the quantity $\Delta$ which for the case of pure states can be algebraically computed in terms of the fidelity. More concretely, let  $\rho=\ket{\psi}\bra{\psi}$ and $\sigma=\ket{\phi}\bra{\phi}$ be pure states, then
\begin{align}
\Delta(\rho,\sigma)=F(\rho,\sigma)-\tr\big( \sqrt{\rho}\sqrt{\sigma}\big)= F(\rho,\sigma)-F^2(\rho,\sigma).
\end{align}
For the same reasons as described for the fidelity, instead of the above pure state expression, we will use the zero temperature limit of~\eqref{eq:deltafid}.

\section{Results}

\subsection{$1$D topological superconductors with exponential decaying hopping
amplitude}
The Kitaev chain~\cite{kit:01} is a one-dimensional model for topological superconductivity, where the hoppings and pairing amplitudes are between nearest neighbours. Various attempts have been made to modify the Kitaev chain by deforming the pairing and hopping amplitudes. We consider the following Hamiltonian with long range couplings~\cite{viy:vod:pup:del:16}

\begin{figure}
\centering
\includegraphics[scale=0.25]
{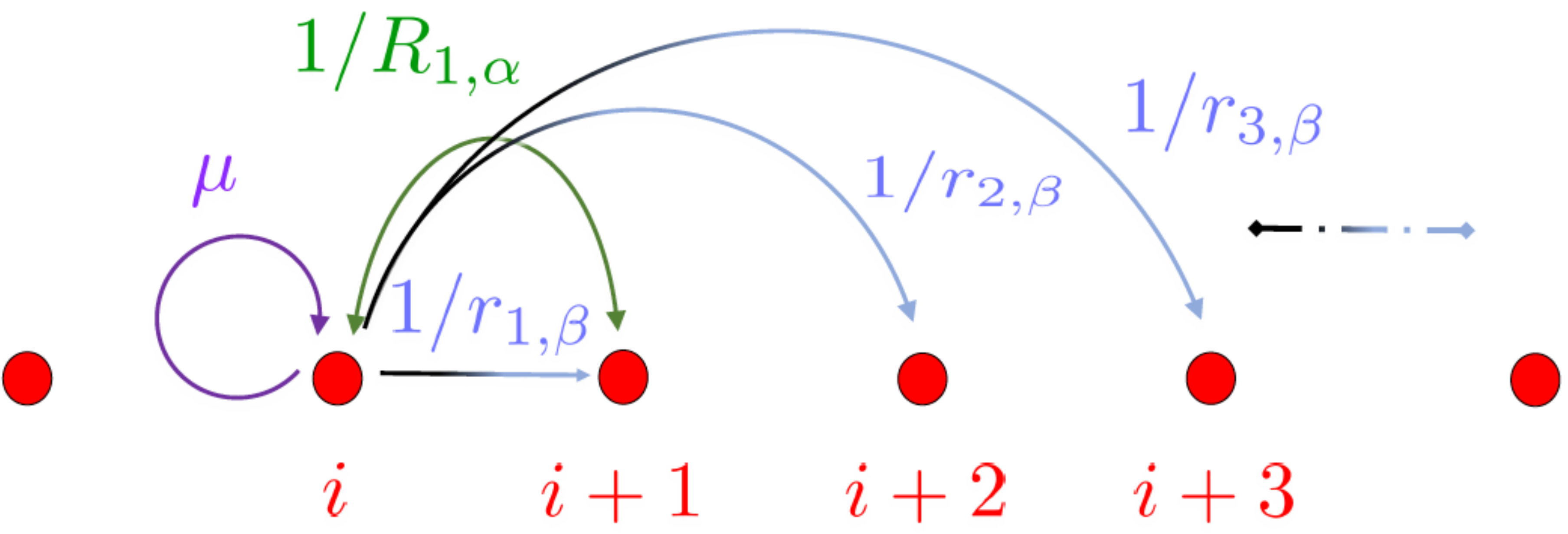}
\caption{Depiction of 1D modified Kitaev superconductor Hamiltonian given in Eq.\eqref{eq:1DTS_longrange} with long-range hopping amplitudes~(blue arrows), strictly short range pairing amplitudes~(green arrow) and chemical potential~(purple arrow).}
 \label{fig:1DLongRangeSC}
\end{figure}

\begin{align}
\mathcal{H} & =\sum_{i=1}^{N}\left(\frac{1}{2}\sum_{l=1}^{N-1}(-\frac{1}{r_{l,\beta}}c_{i+l}^{\dagger}c_{i}+\frac{1}{R_{l,\alpha}}c_{i+l}^{\dagger}c_{i}^{\dagger}+\text{H.c.})-\mu(c_{i}^{\dagger}c_{i}-\frac{1}{2})\right)\label{eq:1DTS_longrange},
\end{align}
where $r_{l,\beta}$ and $R_{l,\alpha}$, control hopping and pairing amplitudes, respectively, and they are functions of the distance $l$, with real parameters $\beta$ and $\alpha$, respectively. The operators $c_{i}^{\dagger}~(c_{i})$ are spinless fermion creation~(annihilation) operators and $\mu$ is the chemical potential. The short range pairing and long range hoppings in $\mathcal{H}$ are illustrated in Fig.~\ref{fig:1DLongRangeSC}. Observe that for strictly nearest-neighbour hopping and pairing amplitudes, i.e., for $1/r_{l,\beta}=2t\delta_{l,1}$ and $1/R_{l,\alpha}=2\Delta\delta_{l,1}$, we recover the original Kitaev chain Hamiltonian
 \begin{align}
\mathcal{H} & =\sum_{i=1}^{N}\left(-t c_{i+1}^{\dagger}c_{i}+\Delta c_{i+1}^{\dagger}c_{i}^{\dagger}+\text{H.c.}-\mu(c_{i}^{\dagger}c_{i}-\frac{1}{2})\right)\label{eq:1DTS_longrange}.
\end{align}

Taking advantage of translational invariance, we can diagonalise the above Hamiltonian by going to Fourier space, obtaining a discrete version of Eq.~\eqref{eq:HBDG},
\begin{align}
\mathcal{H}(q) & =\frac{1}{2N}\sum_{k}\widetilde{\psi}_{k}^{\dagger}H_{\text{BdG}}(k;q)\widetilde{\psi}_{k},\label{eq:FourierSpaceHam}
\end{align}
with $k\in \{\frac{2\pi \ell}{N} : \ell \in \{0,...,N-1\}\}$, $q=(\mu,\beta,\alpha)$, and
\begin{align*}
E(k;q) & =\sqrt{(\mu+g_{\beta}(k))^{2}+f_{\alpha}^{2}(k)},\\
n(k;q) & =-\frac{1}{E(k;q)}(0,f_{\alpha}(k),(\mu+g_{\beta}(k))),
\end{align*}
where 
\[
f_{\alpha}( k)=\sum_{l=1}^{N-1}\frac{\sin( k l)}{R_{l,\alpha}}\ \ \ \ \ \ g_{\beta}( k)=\sum_{l=1}^{N-1}\frac{\cos( kl)}{r_{l,\beta}}.
\]
We study a special case where pairing is considered to be strictly nearest-neighbour, $1/R_{l,\alpha}=\delta_{1,l}$, and hopping terms which decay exponentially of the form $r_{l,\beta}=\exp\big((l-1)/\beta\big)$. This fixes the parameter $\alpha$ and hence we restrict ourselves to the parameters $q\equiv(\mu,\beta)$.

\begin{figure}
\subfloat[\label{fig:1DSuperConductFed}]{\includegraphics[width=2.2in,height=2.3in]{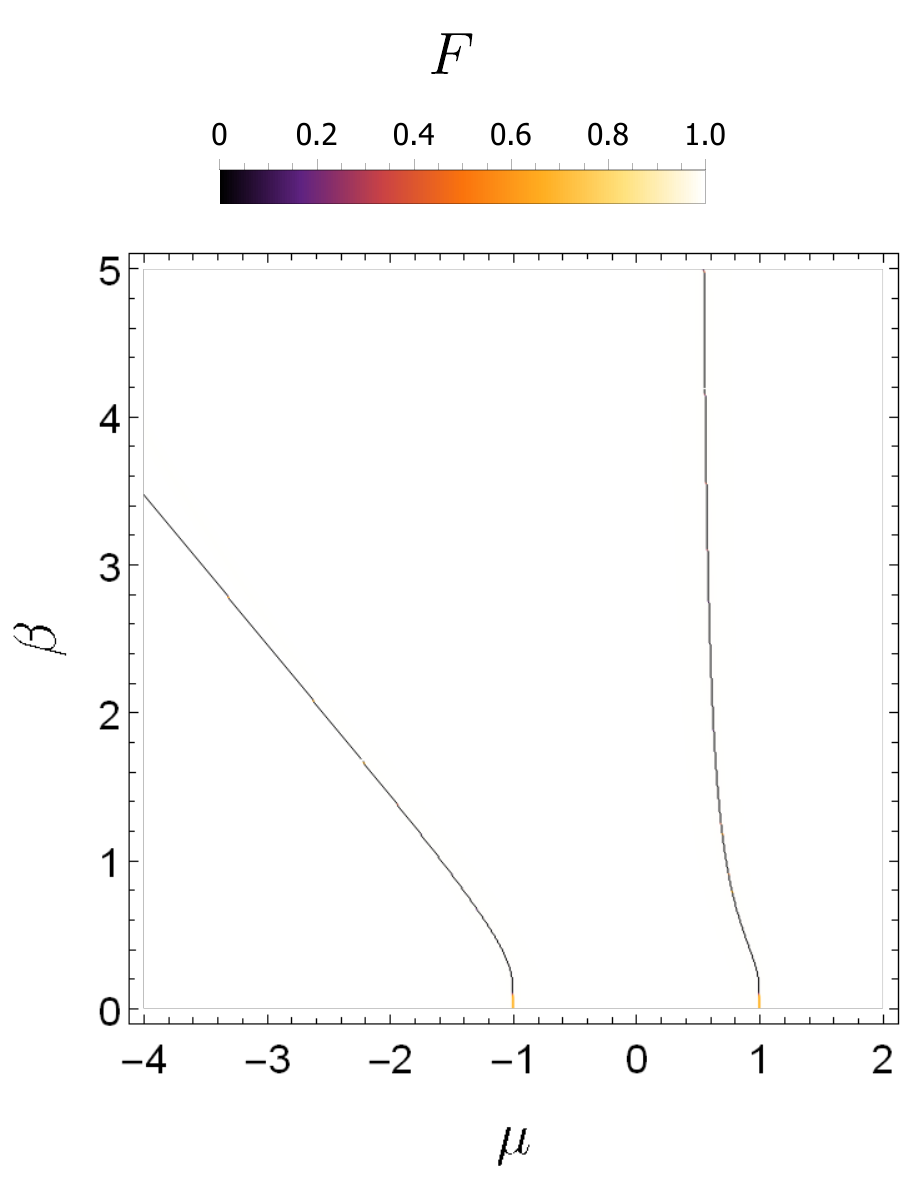}}\subfloat[\label{fig:1DSuperConductUhl}]{\includegraphics[width=2.2in,height=2.3in]{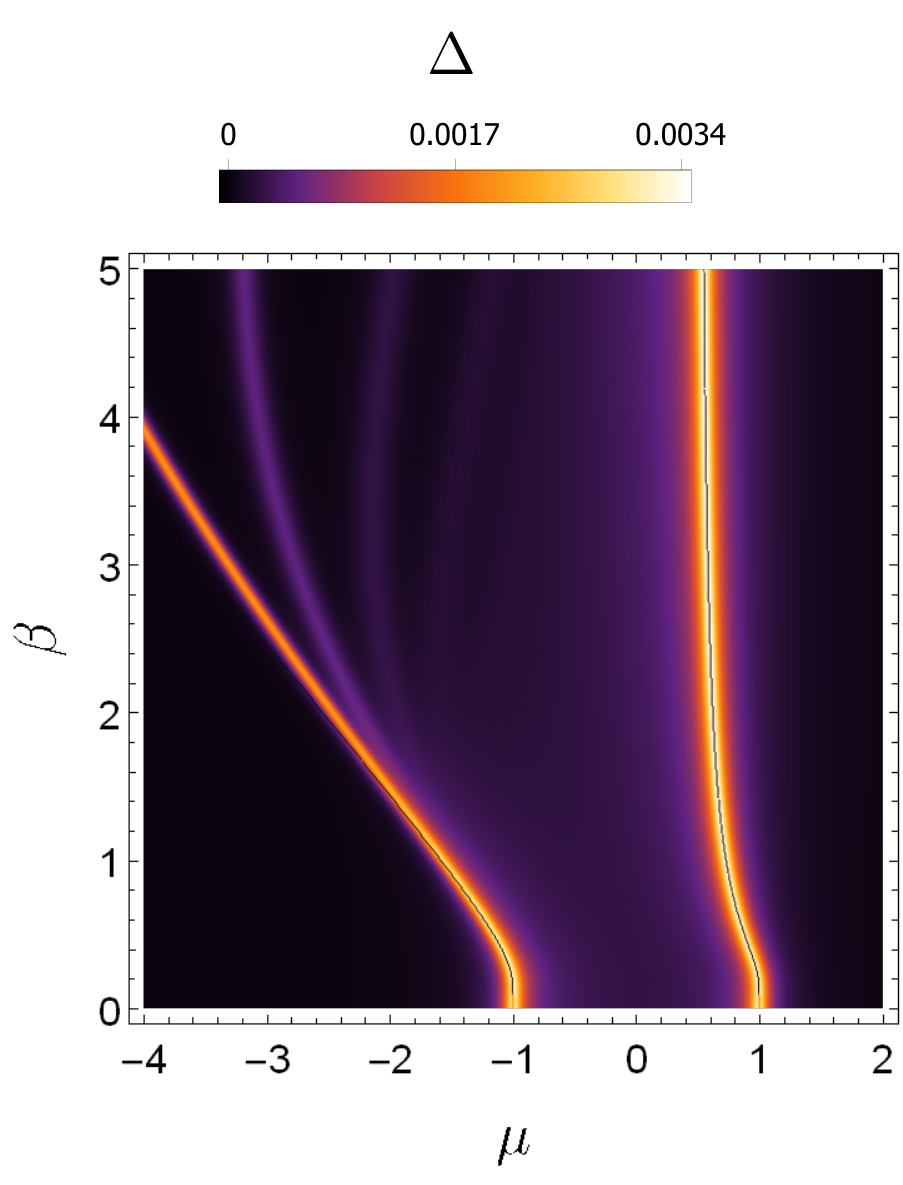}

}\subfloat[\label{fig:1DSuperConductChi}]{\includegraphics[width=2.2in,height=2.3in]{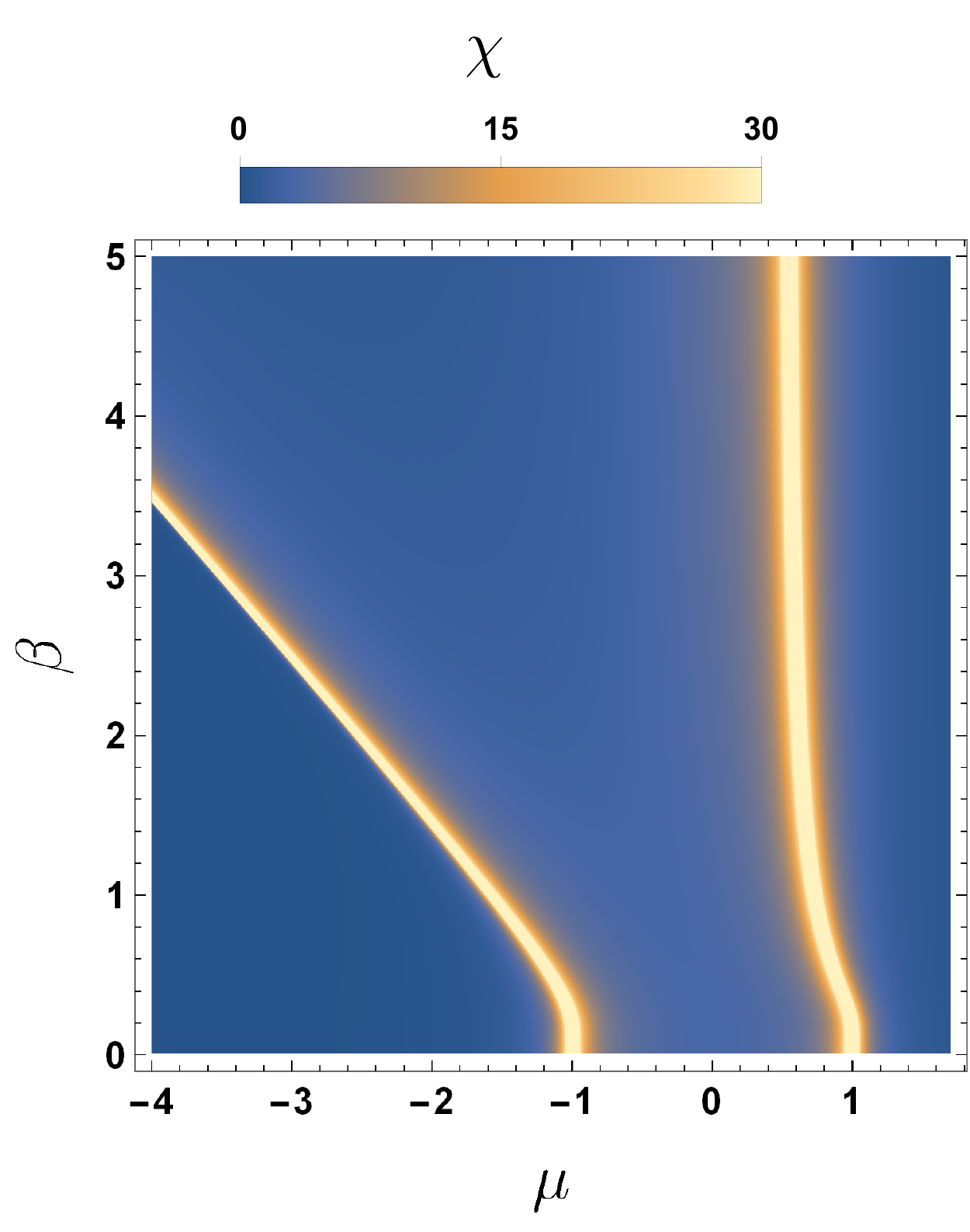}}\caption{
Fidelity $F(\rho,\rho')$ and (b) Uhlmann quantity $\Delta(\rho,\rho')$ for states $\rho=\rho(\mu,\beta)$ and $\rho'=\rho(\mu+\delta\mu,\beta)$ as functions of $(\mu,\beta)$, where $\delta \mu = 0.01$ and $T=10^{-5}$ (c) Ground state fidelity susceptibility $\chi_{\mu\mu}=\chi_{\mu\mu}(\mu,\beta)$ for 1D TSCs as given by Eq.~(\ref{eq:1DTS_longrange}).} \label{fig:1DSuperConduct}

\end{figure}

In the case of a $1$D superconductor with strictly nearest hopping and pairing amplitudes, there is one nontrivial topological phase with winding number $1$ for $-1<\mu<1$, where for other values of $\mu$ the winding number is $0$. On the other hand, when long-range effects are included, i.e., $\beta>0$, the topological phase diagram, given in terms of $\mu$, changes~\cite{viy:vod:pup:del:16}. Our fidelity and $\Delta$ analysis show this exact behaviour, see Fig.~\ref{fig:1DSuperConduct}. We probe the parameter space by considering two nearby points $(\mu, \beta)$ and $(\mu + \delta\mu,\beta)$ and calculate the fidelity $F = F(\rho(\mu,\beta),\rho(\mu + \delta\mu,\beta))$ and $\Delta = \Delta(\rho(\mu,\beta),\rho(\mu + \delta\mu,\beta))$ for $\delta\mu = 0.01$ and temperature $T=10^{-5}$. We refer the reader to the appendices of~\cite{mer:vla:pau:vie:17,ami:mer:val:pau:vie:18} for details of the analytical derivations of the expressions for the fidelity, $\Delta$ and the ground state fidelity susceptibility, also applicable for the $2$D case studied below. Fig.~\ref{fig:1DSuperConductFed} shows the critical lines of topological phase transitions. The value of $\delta \mu$ is chosen such that the two states $\rho(\mu)$ and $\rho(\mu+\delta\mu)$ are generically the same. However, near the critical points, no matter how small $\delta \mu$ is, they are substantially different. Indeed, we can see that $F$ is one everywhere except for the phase transition lines where its value drops down from one. 

Unlike generic models where the fidelity between two different phases becomes zero only in the thermodynamic limit, in the current model, the zero temperature fidelity between two ground states from different phases is exactly zero, no matter how small the parameter difference is. This follows from the fact that at the special momentum points $k=0,\pi$, the vector $n(k)$ is proportional to the $z$ axis and it happens that, for $k=\pi$, it has opposite signs in the different phases leading to a zero factor in the product determining the ground state fidelity. Indeed, when calculating the zero temperature pure state fidelity of a finite system we obtain the {\em exact} zero value in the vicinity of the critical point. Recall that at the critical points, the ground state manifold is degenerate; thus, we avoid such points from this calculation. To observe the exact zero value for fidelity we need to compare states from different phases. To ensure that this will always happen, the parameter difference, say $\delta\mu$, has to be bigger than the spacing between the two consecutive parameter points in which we calculate the ground state fidelity (grid spacing $\Delta\mu$); in our case, $\delta\mu = 0.01 > 0.0099 = \Delta\mu$. We would like to stress that this example shows the generality of the fidelity approach, which allowed us to correctly infer the critical points even without considering physical details of the system.

Fig.~\ref{fig:1DSuperConductUhl} shows the behaviour of $\Delta$: the departure of $\Delta$ from zero shows the nontrivial behaviour along the phase transition lines. Note that the faded curves near the left transition line represent numerical instabilities, as they do not feature in the plots of $F$ and $\chi$. The numerical value used for the number of lattice sites in the calculation of fidelity and the Uhlmann quantity is $N = 80$. In Fig.~\ref{fig:1DSuperConductChi} we plot the $\chi_{\mu\mu}$ which also shows the topological phase transitions. We can see from all the above plots that as $\beta$ goes to zero we recover the usual topological superconductor where criticality exist at $\mu = 1$ and $-1$, respectively. On the other hand, when $\beta$ increases above zero we see the above mentioned drift of the critical points in the parameter space.  

Along with the above analysis, we also performed asymptotic analysis of the ground state fidelity susceptibility in the neighbourhood (both left and right) of the critical points, where we obtain the associated critical exponents. The critical exponents are calculated as done in statistical mechanics  and in the theory of phase transitions, where one wants to understand the nature of the singular point (see for instance~\cite{and:07}). Near a gapless point, only the behaviour near the critical momenta is relevant. Therefore, there should exist some underlying universal behaviour captured in the correlation functions of theory. The fidelity susceptibility can be written in terms of the correlation functions (see Section IV B of~\cite{gu:10}) and therefore would also manifest this universality properties. The power law behaviour is typical for gapless systems, due to generic arguments of conformal invariance.

\begin{figure}[h!]
\par
\subfloat[\label{fig: 1DFidelitySuscepA}$\beta_{c}$ = 1.101, $b = 0.97 \pm 0.0043$]{\includegraphics[width=1.8in,height=1.3in]{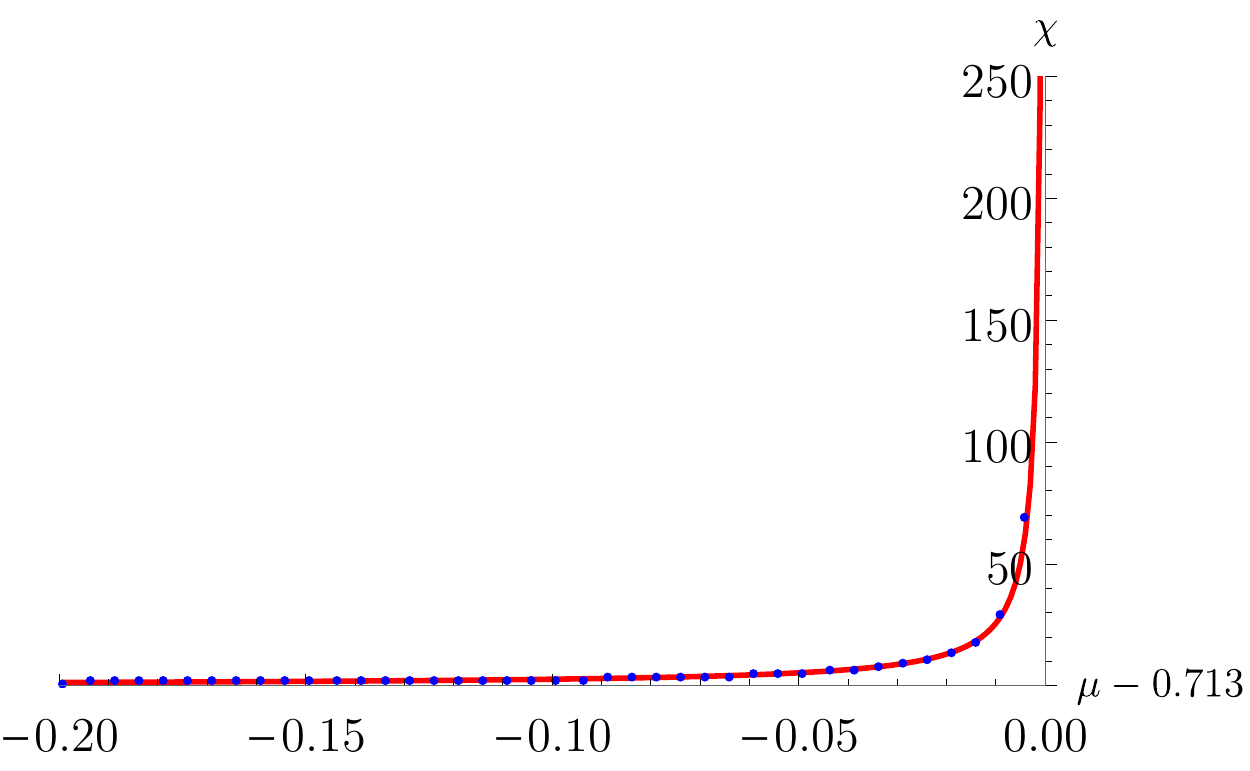}}
\subfloat[\label{fig: 1DFidelitySuscepB} $\beta_{c}$ = 1.101, $b = 0.98 \pm 0.0066$]{\includegraphics[width=1.8in,height=1.3in]{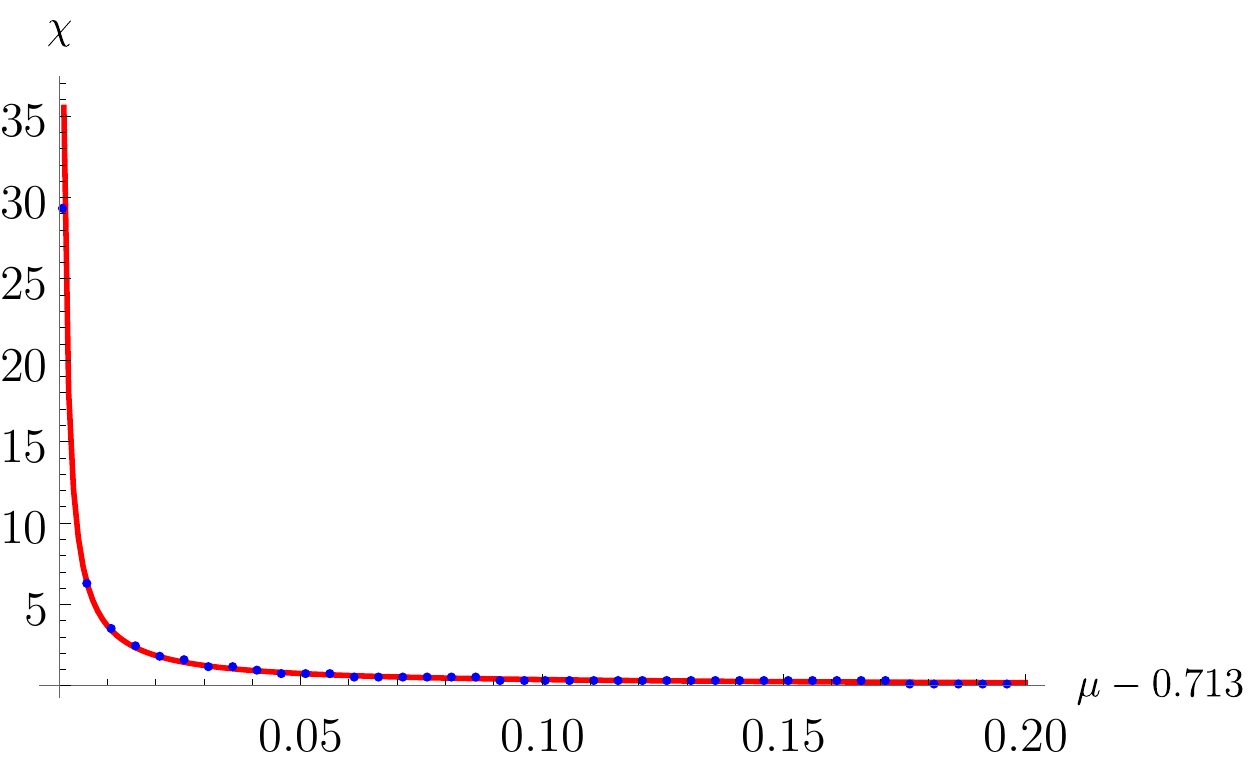}}
\subfloat[\label{fig: 1DFidelitySuscepC}$\beta_{c}$ = 2.101, $b = 0.95 \pm 0.0047$]{\includegraphics[width=1.8in,height=1.3in]{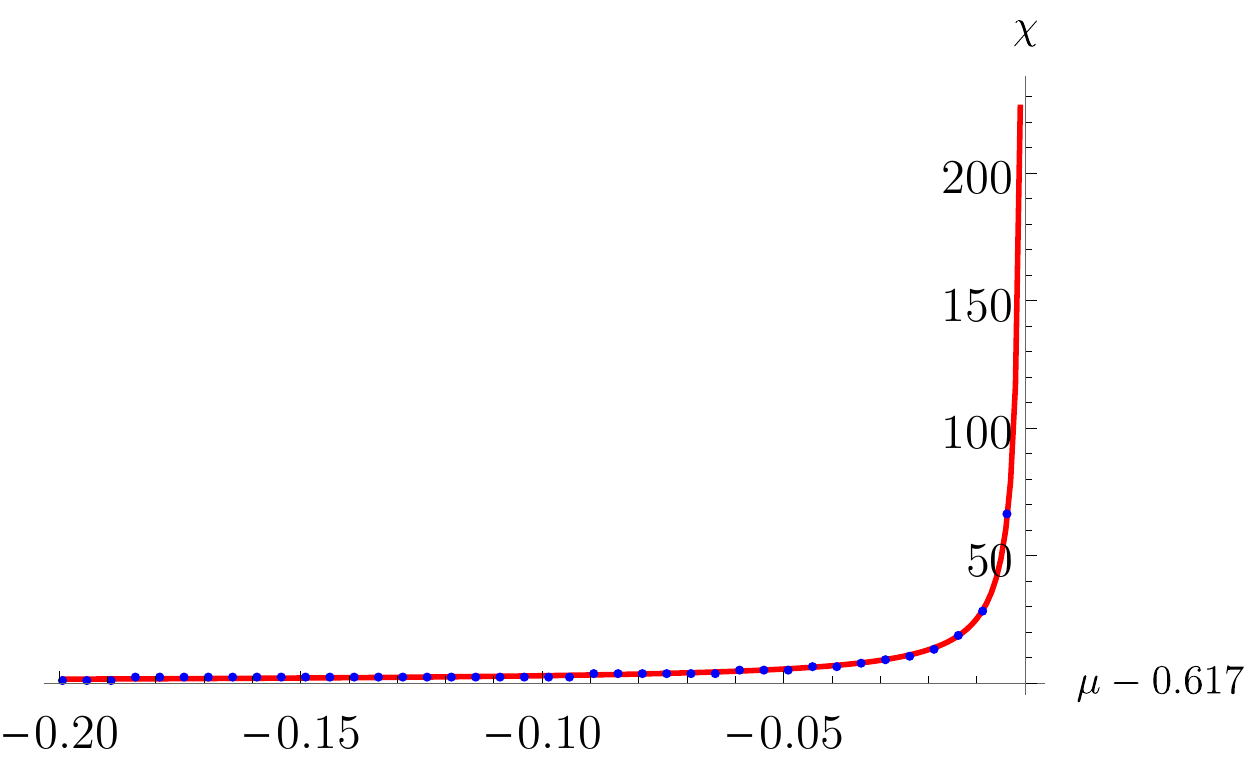}}
\subfloat[\label{fig: 1DFidelitySuscepD}$\beta_{c}$ = 2.101, $b = 0.97 \pm 0.0026$]{\includegraphics[width=1.8in,height=1.3in]{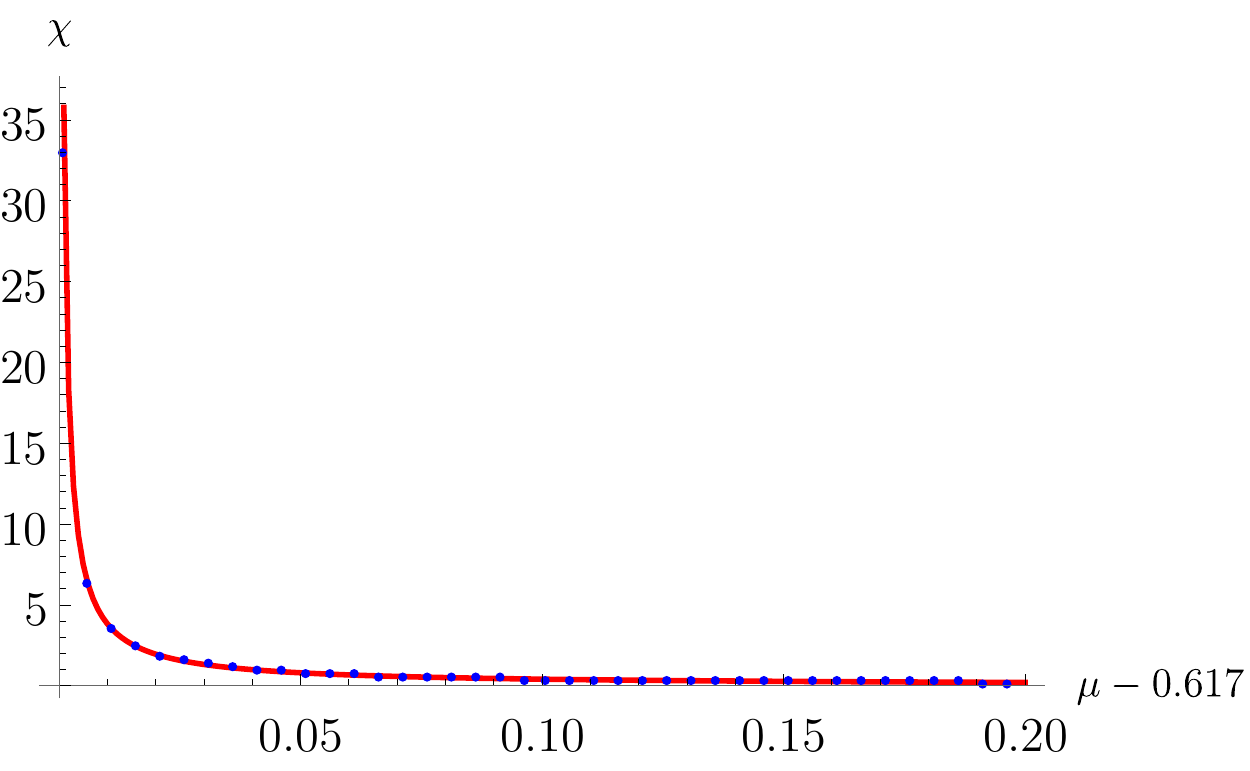}}

\par

\subfloat[\label{fig: 1DFidelitySuscepE} $\beta_{c}$ = 1.101, $b = 1.98 \pm 0.0039$  ]{\includegraphics[width=1.8in,height=1.3in]{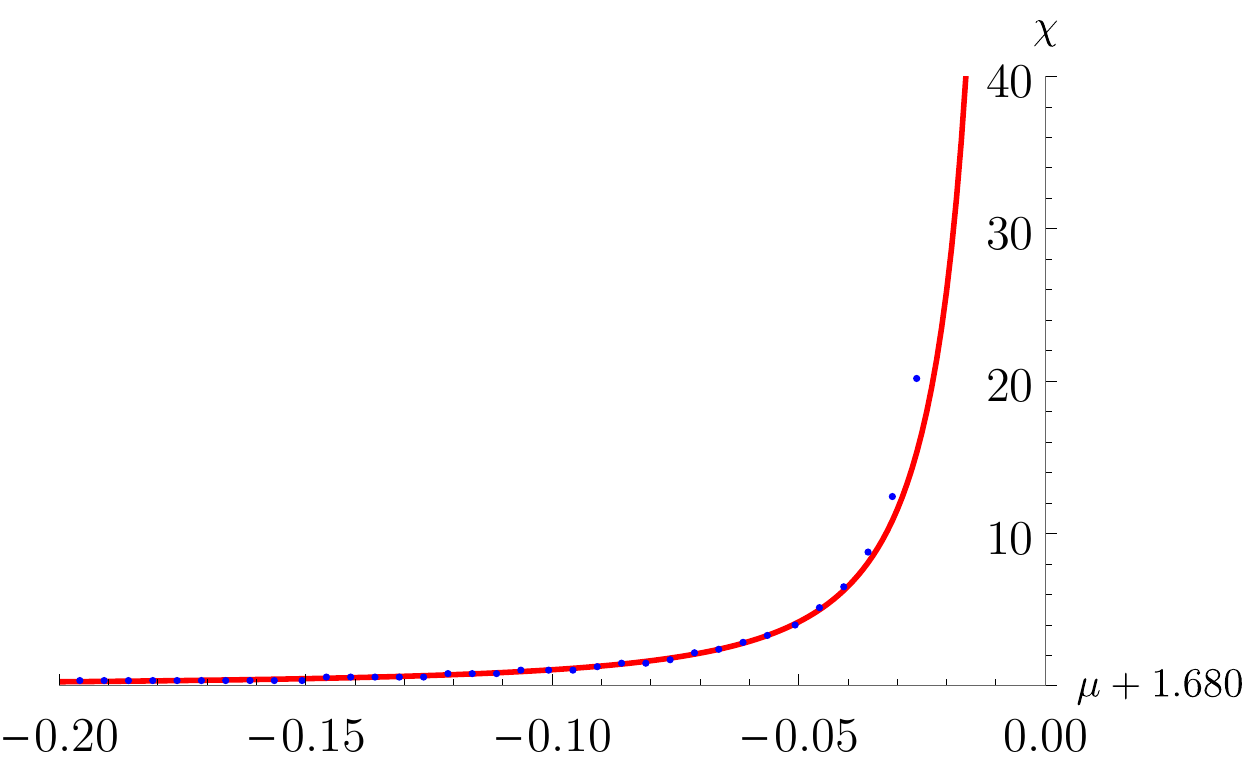}}
\subfloat[\label{fig: 1DFidelitySuscepF} $\beta_{c}$ = 1.101, $b = 1.01 \pm 0.0696$]{\includegraphics[width=1.8in,height=1.3in]{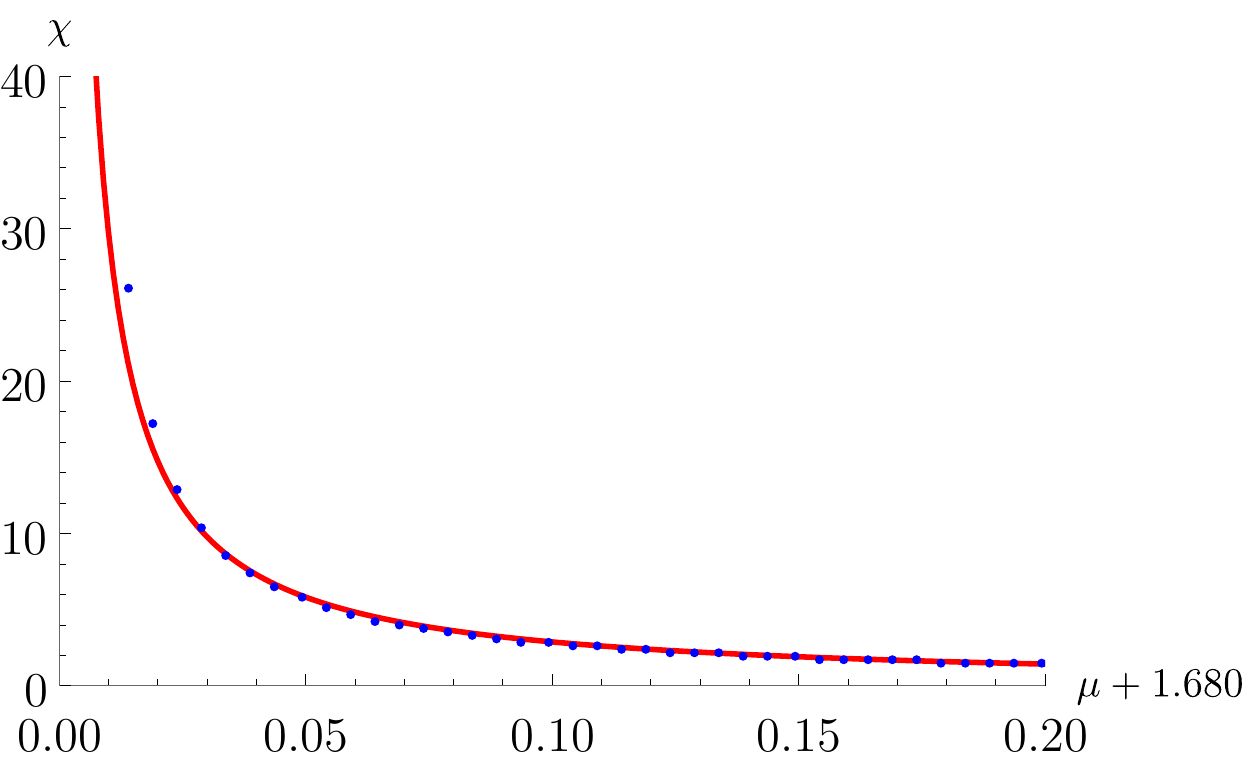}}
\subfloat[\label{fig: 1DFidelitySuscepG}$\beta_{c}$ = 2.101, $b = 1.90 \pm 0.0239$]{\includegraphics[width=1.8in,height=1.3in]{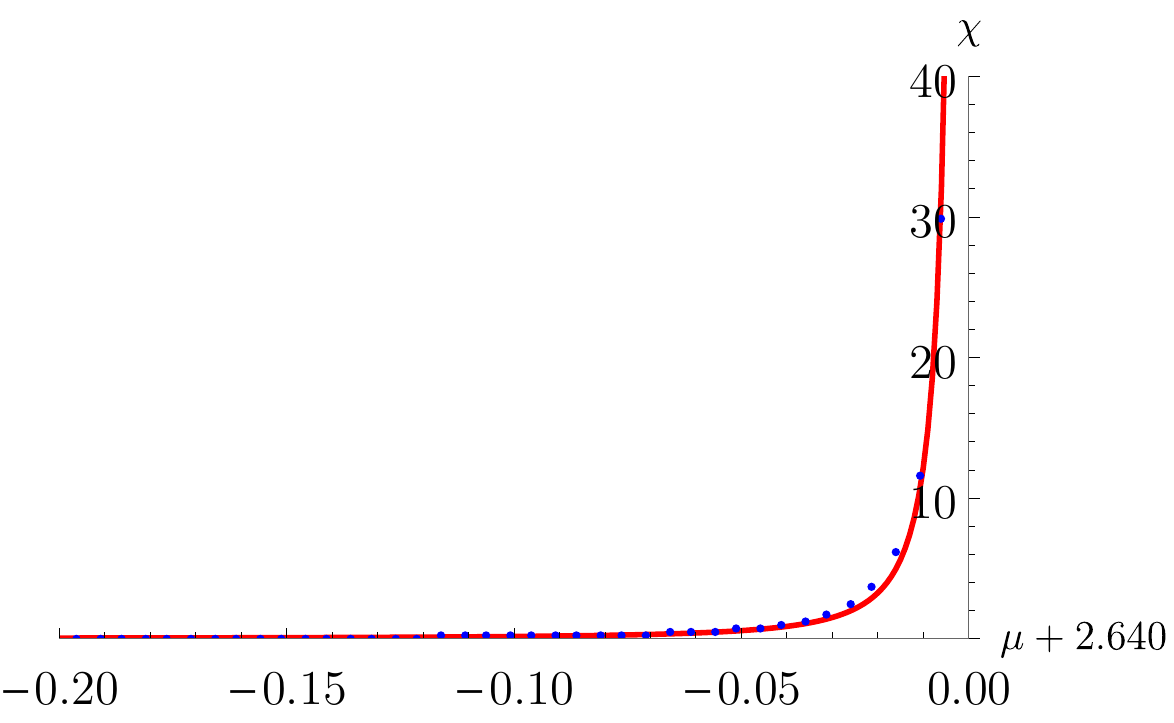}}
\subfloat[\label{fig: 1DFidelitySuscepH}$\beta_{c}$ = 2.101, $b = 1.02 \pm 0.0147$]{\includegraphics[width=1.8in,height=1.3in]{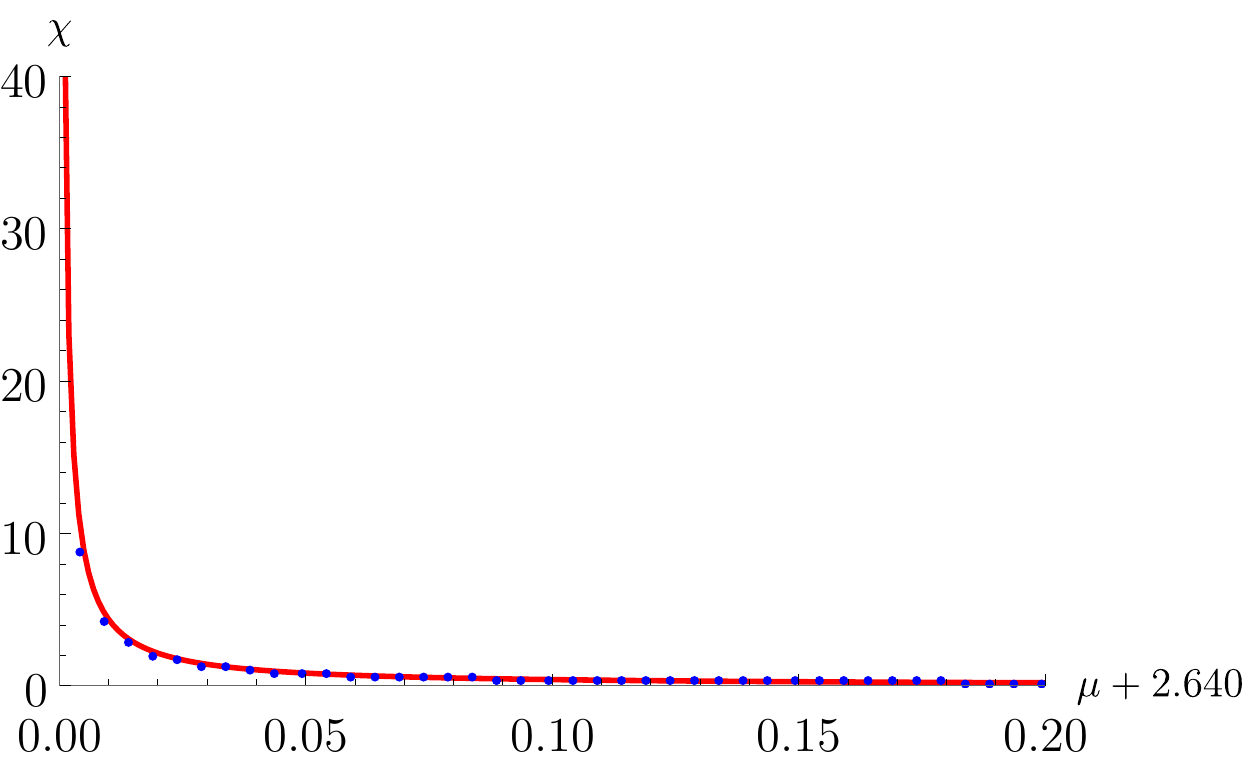}}
\par

\caption{The ground state fidelity susceptibility $\chi_{\mu\mu}$ as a function of $(\mu-\mu_c)$ with fixed values of $\beta=\beta_{c} = 1.101$ for (a), (b), (e),  (f) and $\beta=\beta_{c} = 2.101$ for (c), (d), (e), (f)   in the neighbourhood of different critical $\mu_c$: (a) and (b) $\mu_c =0.713$, (c) and (d) $\mu_c =0.617$, (e) and (f) $\mu_c =-1.68$, (g) and (h) $\mu_c =-2.64$. The points and red curve are the result of numerical integration and of the fit respectively.}
\label{fig: 1DFidelitySuscep}
\end{figure}

 A least squares fit of the susceptibility to a power law $\chi\propto|\mu-\mu_c|^{-b}$ in a neighbourhood of $\mu_c$ was performed for different critical $ \beta_c$, as shown in Fig.~\ref{fig: 1DFidelitySuscep}. In obtaining the parameter $b$ and the associated uncertainty, we used the standard formulas from linear regression (see, for instance,~\cite{lan:heb:gue:osh:zim:17}, p.~462) on a $\log$-$\log$ plot. For $\beta_c = 1.101$, and we found:
\begin{itemize}
\item [i)] in the neighborhood of $\mu_c = 0.713$, exponents $b = 0.97 \pm 0.0043$ from the left, see Fig.~\ref{fig: 1DFidelitySuscepA}, and $b = 0.98 \pm 0.0066$ from the  right, see Fig.~\ref{fig: 1DFidelitySuscepB}.
\item[ii)] in the neighborhood of $\mu_c = -1.680$, exponents $b = 1.98 \pm 0.0039$ from the left, see Fig.~\ref{fig: 1DFidelitySuscepE} and $b = 1.01 \pm 0.0696$ from the right, see Fig.~\ref{fig: 1DFidelitySuscepF}. 
\end{itemize}
Similarly, for $\beta_c = 2.101$, we found:
\begin{itemize}
\item[i)] in the neighbourhood of $\mu_c = 0.617$, exponents $b = 0.95 \pm 0.0047$ from the left, see Fig.~\ref{fig: 1DFidelitySuscepC}, and $b = 0.97 \pm 0.0026$ from the right, see Fig.~\ref{fig: 1DFidelitySuscepD}.
\item[ii)] in the neighbourhood of $\mu_c = -2.640$, exponents $b = 1.90 \pm 0.0339$ from the left, see Fig.~\ref{fig: 1DFidelitySuscepG}, and $b = 1.02 \pm 0.0147$ from the right, see Fig.~\ref{fig: 1DFidelitySuscepH}.
\end{itemize}

\subsection{$2$D topological superconductors with long-range effects}
In the same way as in case of 1D TSCs, modifications are made in the hopping and pairing amplitudes of short range 2D chiral superconductors. We consider the following modified Hamiltonian model~\cite{viy:lia:del:ang:18}
\begin{align}
\mathcal{H}=\sum_{m,l}\left(\sideset{}{'}\sum_{r,s} (-\frac{t}{d_{r,s}^{\beta}}c_{l+r,m+s}^{\dagger}c_{l,m}+\frac{\Delta}{d_{r,s}^{\alpha+1}}(r+\mathrm{i}s)c_{l+r,m+s}^{\dagger}c_{l,m}^{\dagger})-(\mu-4t)c_{l,m}^{\dagger}c_{l,m}\right)+\text{H.c.}, 
\label{eq:2DTS_longrange}
\end{align}	
where $\sum_{r,s}^{'}$ means that $r,s=0$ are excluded from the sum. Again the $c_i^{\dagger}$'s ($c_i$'s) are spinless fermion creation~(annihilation) operators. The hopping amplitude decays as $1/d^\beta_{r,s}$ and the pairing amplitude as $1/d_{r,s}^{\alpha}$, where $d_{r,s} =\sqrt{r^2+s^2}$ is the Euclidean distance between the coupled sites. One can recover the usual short range 2D chiral p-wave superconductor Hamiltonian from above model by restricting amplitudes to only nearest neighbours. Observe that the $r$ and $s$ indices control the hoppings and pairings along the $x$ and $y$ directions, respectively. If we restrict hopping and pairing amplitudes to nearest neighbours only, this would be equivalent to consider only $(r,s)\in \{(1,0),(0,1)\}$. This yields
\begin{align}
\mathcal{H}=\sum_{l,m}\left( (-\frac{t}{d_{1,0}^{\beta}}c_{l+1,m}^{\dagger}c_{l,m}-\frac{t}{d_{0,1}^{\beta}}c_{l,m+1}^{\dagger}c_{l,m}+\frac{\Delta}{d_{1,0}^{\alpha+1}}c_{l+1,m}^{\dagger}c_{l,m}^{\dagger}+\mathrm{i}\frac{\Delta}{d_{0,1}^{\alpha+1}}c_{l,m+1}^{\dagger}c_{l,m}^{\dagger})-(\mu-4t)c_{l,m}^{\dagger}c_{l,m}\right)+\text{H.c.}
\label{eq:2DTS_longrange}
\end{align}
Setting $d_{0,1}=d_{1,0}=1$, we recover the well known short range chiral p-wave superconductors~\cite{ber:hug:13,iva:01,rea:gre:00,vol:99} described by the following Hamiltonian
\begin{align}
\mathcal{H}=	\sum_{l,m}\left( -t( c_{l+1,m}^{\dagger}c_{l,m}+ c_{l,m+1}^{\dagger}c_{l,m})+\Delta (c_{l+1,m}^{\dagger}c_{l,m}^{\dagger}+\mathrm{i} c_{l,m+1}^{\dagger}c_{l,m}^{\dagger})-(\mu-4t)c_{l,m}^{\dagger}c_{l,m}\right)+\text{H.c.}
\end{align}
Note that the same Hamiltonian can be obtained by taking $\alpha$ and $\beta$ to $+\infty$. In this limit only the terms $d_{1,0}=d_{0,1}=1$ from the primed sum survive, and the usual short-range superconductors are obtained \cite{viy:lia:del:ang:18}. Here onward we take $\Delta = t = 1/2$. Considering periodic boundary conditions, i.e., putting our system on a torus, the Hamiltonian in Eq. \eqref{eq:2DTS_longrange} can be brought into the form given in Eq.~\eqref{eq:HBDG} with
\begin{align}
H_{\text{BdG}}(\mathbf{k})=f_{\alpha}(\mathbf{k})\sigma_{x}+h_{\alpha}(\mathbf{k})\sigma_{y}-2\big[\mu-4t+tg_{\beta}(\mathbf{k})\big]\sigma_{z},
\end{align}

where
\begin{align}
f_{\alpha}(\mathbf{k})=\sideset{}{'}\sum_{r,s}\frac{s\sin(k_{x}r+k_{y}s)}{d_{r,s}^{\alpha+1}}, \
 h_{\alpha}(\mathbf{k})=\sideset{}{'}\sum_{r,s}\frac{r\sin(k_{x}r+k_{y}s)}{d_{r,s}^{\alpha+1}},\  g_{\beta}(\mathbf{k})&=\sideset{}{'}\sum_{r,s}\frac{\cos(k_{x}r+k_{y}s)}{d_{r,s}^{\beta}},
\end{align}
and $\bf {k} = (k_x, k_y)$. It is known~\cite{viy:lia:del:ang:18} that in the case of the short-range $2$D TSC described above, in the limit $\alpha,\beta\to +\infty$, there exist two non-trivial topological phases with Chern numbers $C= 1$, for $0 < \mu < 2 $, and $C= -1$, for $2 < \mu < 4$. The points of quantum phase transition are given by $\mu_c = 0, 2, 4$. Switching on the long-range effects, the critical points start drifting, and as a result the nontrivial topological phase with $C = -1$ gets suppressed, while the $C = 1$ phase gets enlarged. 

\begin{figure}[h!]
\subfloat[\label{fig:2DSuperConductFid}]{\includegraphics[scale=0.276]{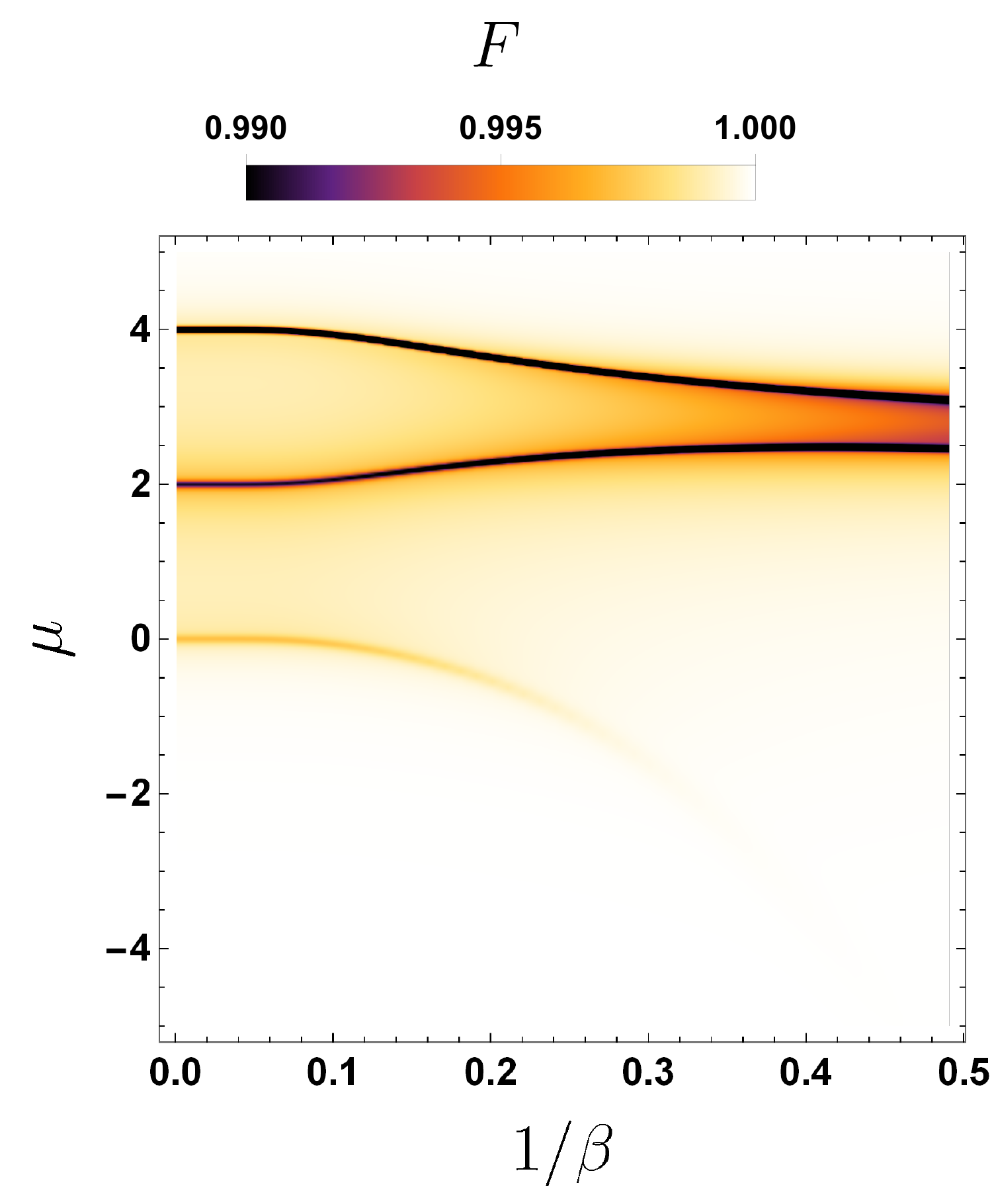}}\subfloat[\label{fig:2DSuperConductUhl}]{\includegraphics[scale=0.276]{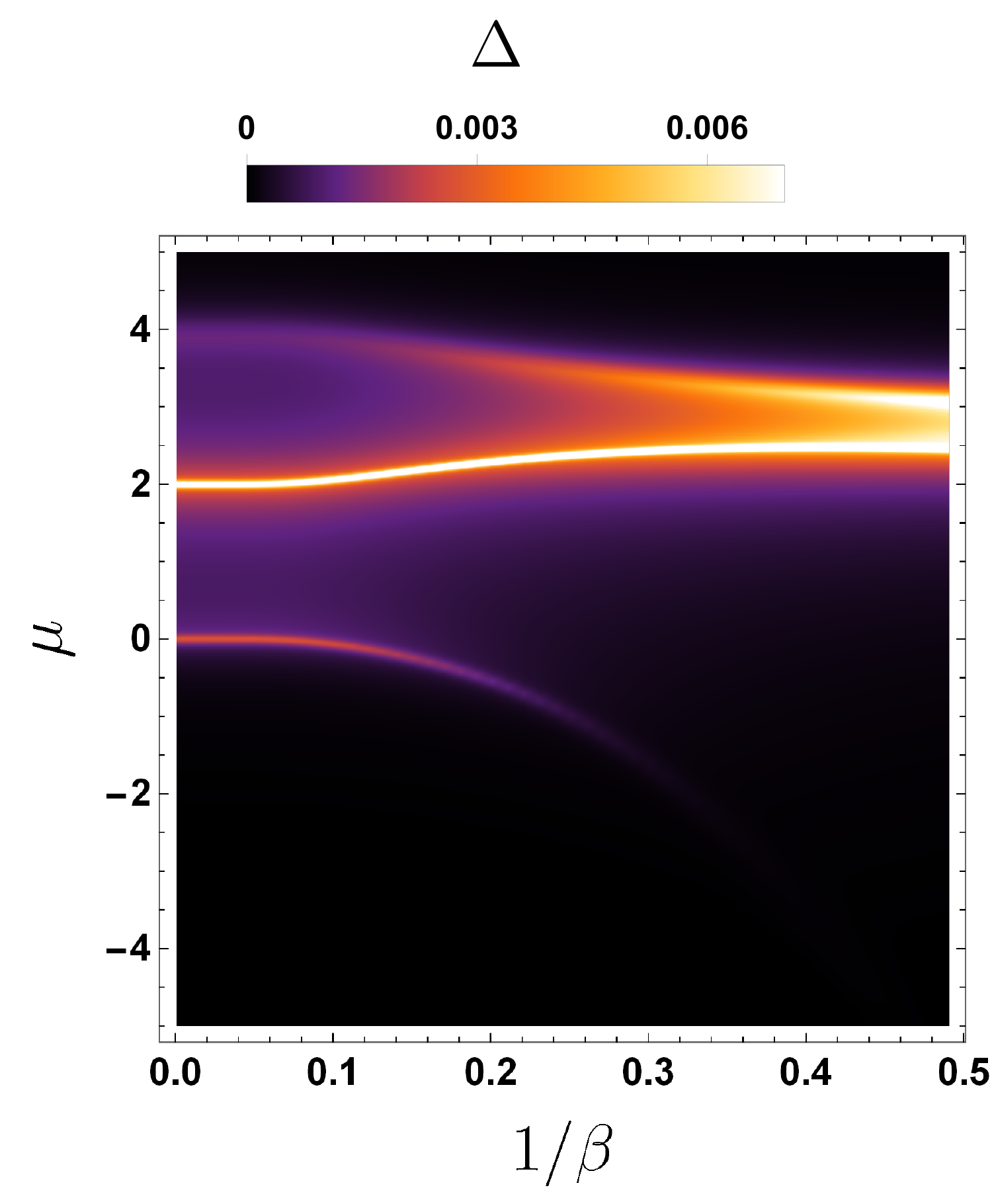}

}\subfloat[\label{fig:2DSuperConductSuc}]{\includegraphics[scale=0.33]{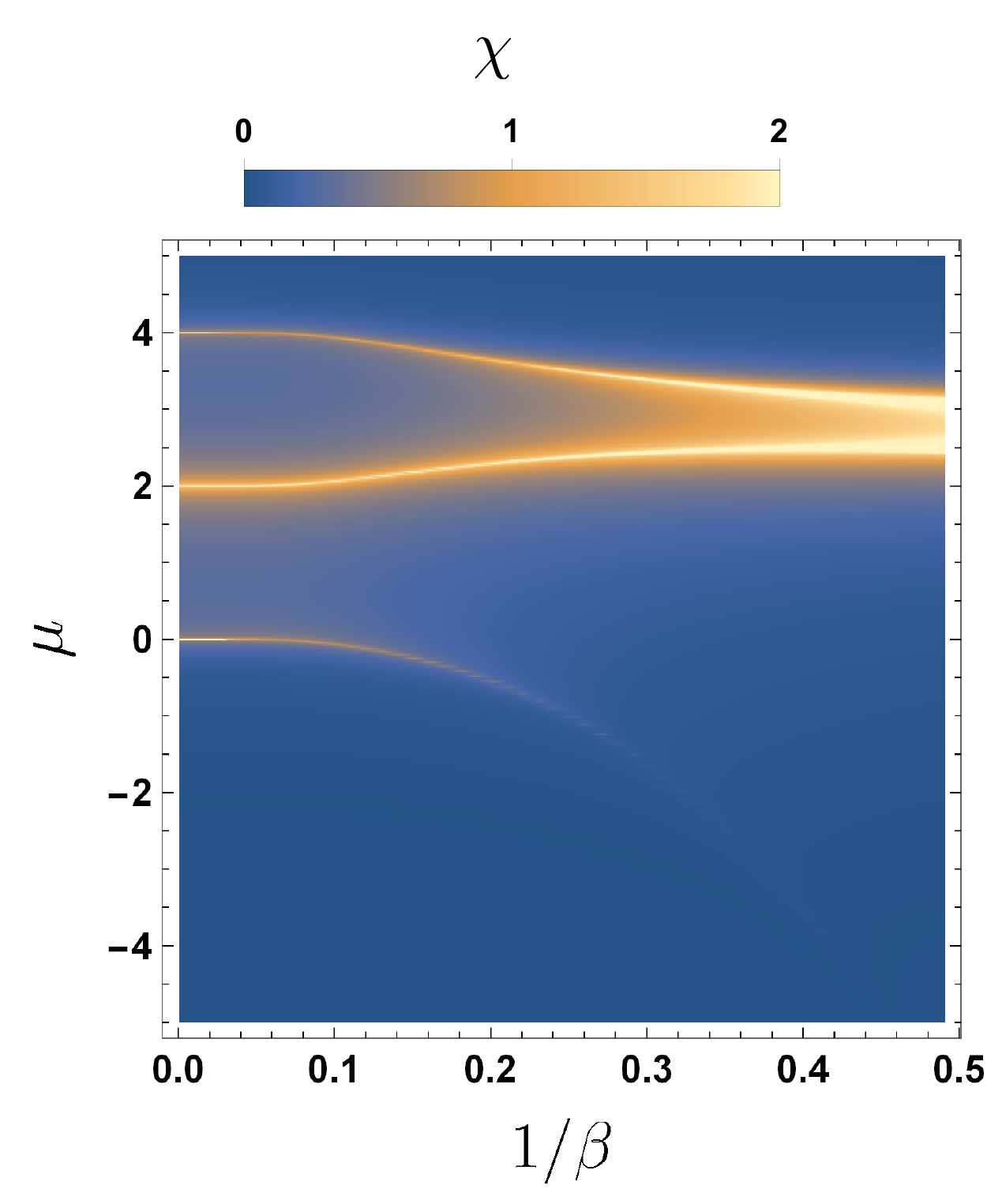}}

\caption{
(a) Fidelity $F(\rho,\rho')$ and (b) Uhlmann quantity $\Delta(\rho,\rho')$ for states $\rho=\rho(\mu,\beta)$ and $\rho'=\rho(\mu+\delta\mu,\beta)$ as functions of $(\mu,\beta)$, where $\delta \mu = 0.01$ and $T=6\times 10^{-3}$. Note that, just like in the 1D case, if we were to choose smaller temperatures and finer grid spacings, we would have observed more prominent behaviour of $F$ and $\Delta$ along the lines of phase transitions. Due to higher complexity of the 2D problem and the rather limited computational power available to us, we omitted such finer analysis. (c) Ground state fidelity susceptibility $\chi_{\mu\mu}=\chi_{\mu\mu}(\mu,\beta)$ for 2D TSCs as given by Eq.~(\ref{eq:2DTS_longrange}).}
\label{fig:2DSuperConduct}
\end{figure}

The plots for the fidelity, $\Delta$ and the ground state fidelity susceptibility $\chi_{\mu\mu}$ are shown in Fig.~\ref{fig:2DSuperConduct}, where we take $\beta = \alpha$. The fidelity and $\Delta$ are obtained between two close states $\rho(\mu, \beta)$ and $\rho(\mu + \delta\mu, \beta)$ for $\delta\mu = 0.01$, temperature $T=6\times 10^{-3}$ and grid spacing $\Delta \mu =0.02$. In numerical calculations of the fidelity and $\Delta$, the number of sites along the $x$~($y$) direction is taken to be $N_x =16$~($N_y =16$). The departure of fidelity from $1$ and $\Delta$ from $0$ signals the quantum phase transition lines, as shown in Fig.~\ref{fig:2DSuperConductFid} and Fig.~\ref{fig:2DSuperConductUhl}, respectively. The ground state fidelity susceptibility also shows the phase transition lines in the parameter space, see~Fig.~\ref{fig:2DSuperConductSuc}. When calculating the zero temperature pure state fidelity of a finite system we obtain, as in the 1D case, the exact zero value in the vicinity of the critical point. The reason is because the Bogoliubov eigenmodes in different phases are not adiabatically connected. Therefore there exists at least one momentum for which the overlap between these modes is zero.

\begin{figure}[h!]
\par
\subfloat[\label{fig: 2DFidelitySuscepA}$1/\beta_{c}$ = 0.0181, $b = 0.19 \pm 0.0014$]{\includegraphics[width=1.8in,height=1.3in]{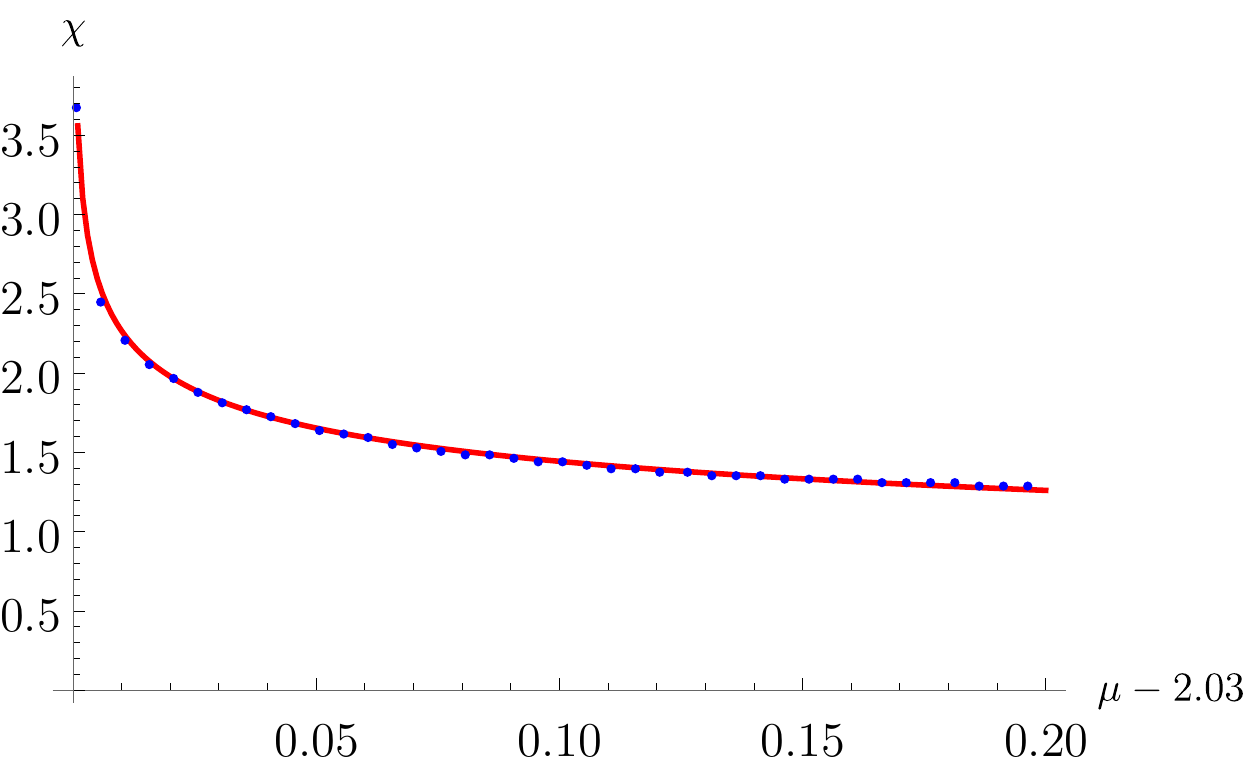}}
\subfloat[\label{fig: 2DFidelitySuscepB}$1/\beta_{c}$ = 0.1223, $b = 0.17 \pm 0.0050$]{\includegraphics[width=1.8in,height=1.3in]{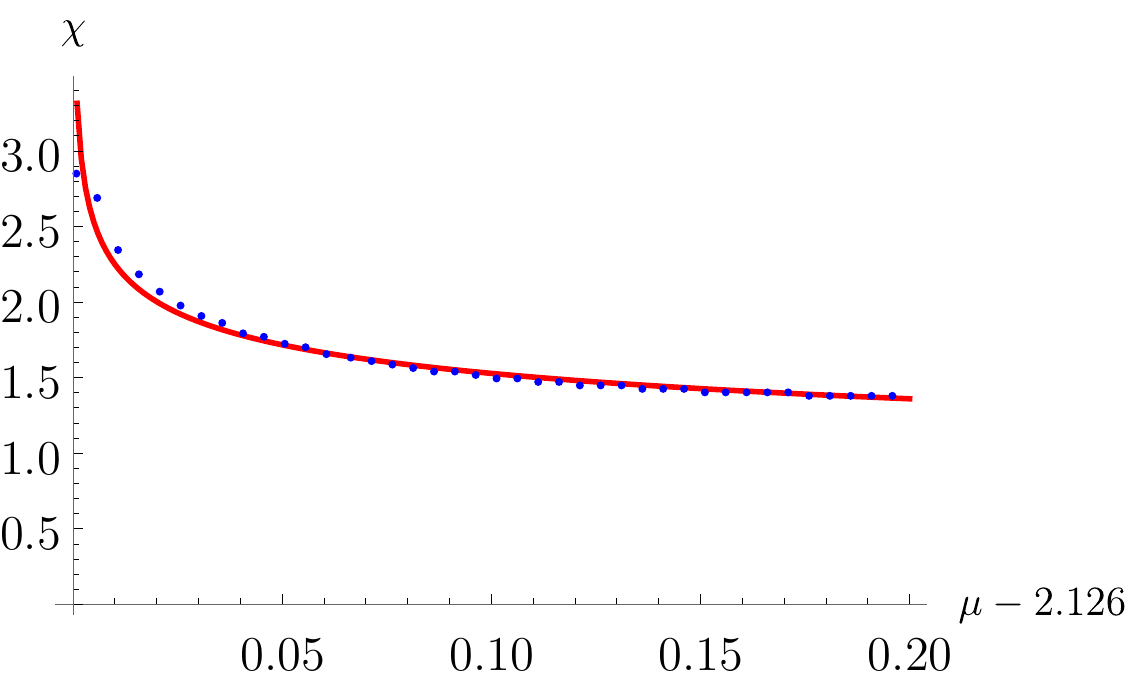}}
\subfloat[\label{fig: 2DFidelitySuscepC}$1/\beta_{c}$ = 0.0181 $b = 0.96 \pm 0.0236$]{\includegraphics[width=1.8in,height=1.3in]{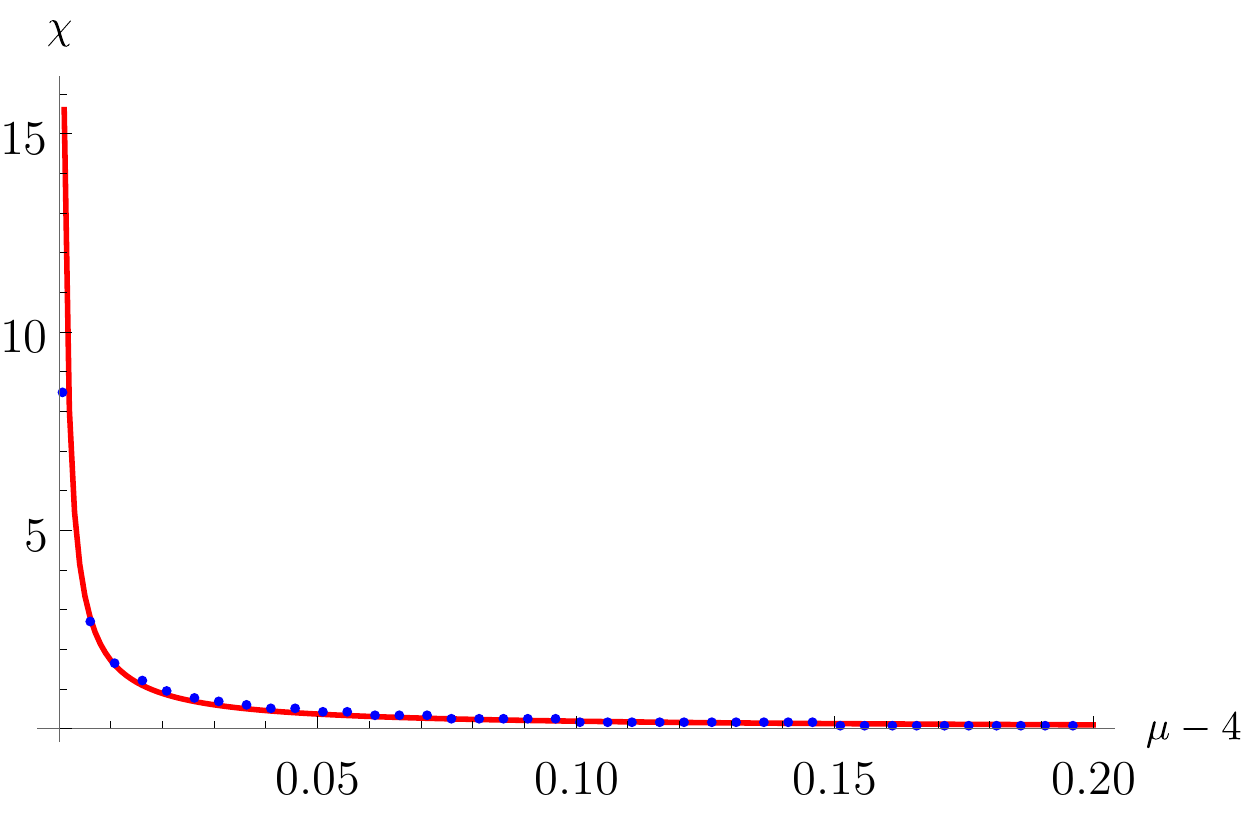}}
\subfloat[\label{fig: 2DFidelitySuscepD}$1/\beta_{c}$ = 0.1223, $b = 0.79 \pm 0.0370$]{\includegraphics[width=1.8in,height=1.3in]{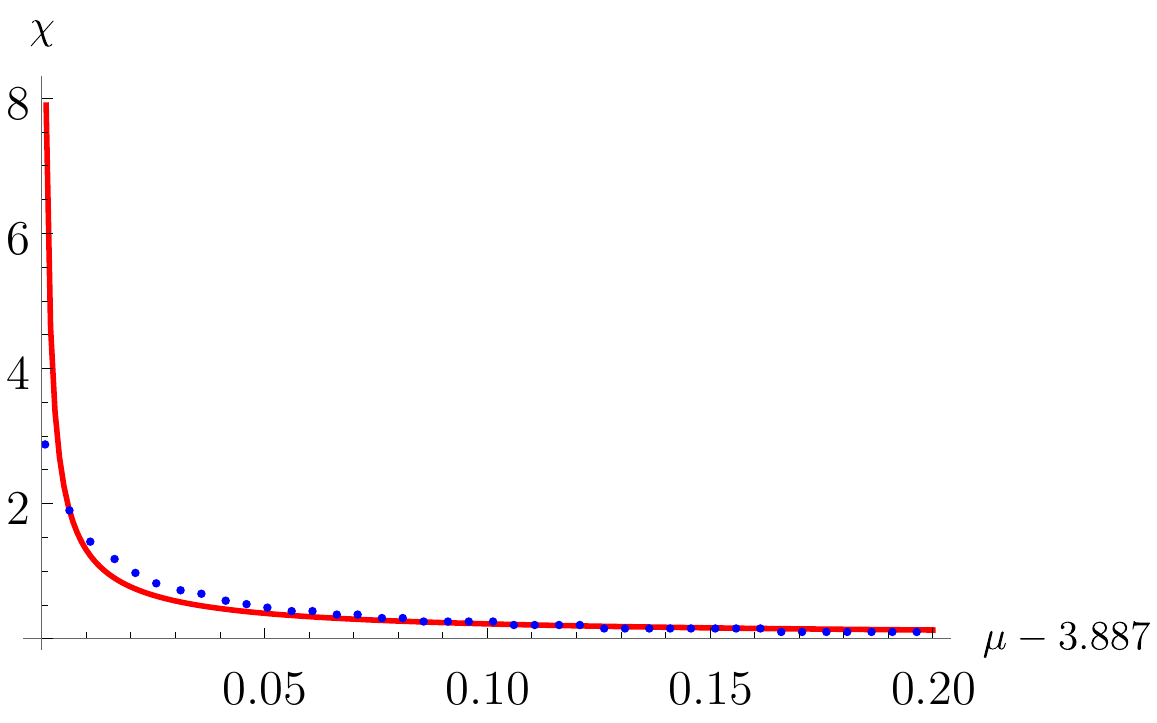}}

\par
\caption{The ground state fidelity susceptibility $\chi_{\mu\mu}$ as a function of $\mu-\mu_c$ with fixed values of $1/\beta=1/\beta_c = 0.0181$ for (a), (c) and $1/\beta=1/\beta_c = 0.1223$ for (b), (d) in the neighbourhood of different critical $\mu_{c}$:(a) $\mu_{c}=2.03$, (b) $\mu_{c}=2.126$, (c) $\mu_{c}=4$ and (d) $\mu_{c}=3.887$.}
\label{fig: 2DFidelitySuscep}
\end{figure}

We also performed the asymptotic analysis of the ground state fidelity susceptibility in the neighbourhood of critical $\mu_c$ for fixed value of $\beta=\beta_c$ as shown in Fig. (\ref{fig: 2DFidelitySuscep}). For $1/\beta_c = 0.0181$, we performed a least squares fit of the susceptibility to a power law $\chi\propto|\mu-\mu_c|^{-b}$, yielding:
\newpage
\begin{itemize}
\item [i)] in the neighbourhood of $\mu_c = 2.03$, exponent $b = 0.19 \pm 0.0014$ from the right, see Fig.~\ref{fig: 2DFidelitySuscepA}.
\item[ii)] in the neighbourhood of $\mu_c = 4.00$, exponent $b = 0.96 \pm 0.0181$ from the right, see Fig.~\ref{fig: 2DFidelitySuscepC}. 
\end{itemize}
Similarly, for $1/\beta_c = 0.1223$, the least squares fit yields:
\begin{itemize}
\item [i)] in the neighbourhood of $\mu_c = 2.13$, exponent $b = 0.17 \pm 0.0050$ from the right, see Fig.~\ref{fig: 2DFidelitySuscepB}.
\item[ii)] in the neighbourhood of $\mu_c = 3.89$, exponent $b = 0.79 \pm 0.0370$ from the right, see Fig.~\ref{fig: 2DFidelitySuscepD}. 
\end{itemize}




\section{Conclusions}
In this work we performed an information geometric analysis of $1$D and $2$D topological superconductors with long-range hopping and pairing amplitudes. The fidelity, $\Delta$ and the ground state fidelity susceptibility confirmed the phase diagram as obtained in the recent works~\cite{viy:vod:pup:del:16, viy:lia:del:ang:18}. We also performed the analysis of the critical exponents for the ground state fidelity susceptibility. Our analysis in terms of fidelity allows for a natural extension of the study to finite temperatures, where the notion of topological phase is yet to be completely understood. Moreover, it would be interesting to understand the interplay between long range phenomena and temperature fluctuations.
Traditionally, the impact of thermal fluctuations is often studied in terms of local order parameters. In the case of topological phase transitions, the role of order parameters could be assigned to global topological invariants, but they are applicable only at the zero temperature. Thus, finding possible observable quantities that could probe topological orders at finite temperatures is an important open problem. The fidelity approach offers one such candidate: an {\em optimal observable} for the quantum fidelity. In short, measuring an optimal observable on the systems in states $\rho$ and $\sigma$ produces two classical probability distributions whose classical fidelity is equal to the quantum fidelity between $\rho$ and $\sigma$ (for details, see~\cite{nie:chu:02}, Section 9.2.2, page 412). We believe analysing such possibility could possibly be an intriguing line of future research.

\acknowledgements{

STA, BM and NP acknowledge the support of SQIG -- Security and Quantum Information Group, under the Funda\c{c}\~ao para a Ci\^{e}ncia e a Tecnologia (FCT) project UID/EEA/50008/2019, and European funds, namely H2020 project SPARTA.
STA acknowledges the support from DP-PMI and FCT (Portugal) through the grant PD/ BD/113651/2015. BM and NP acknowledge projects QuantMining POCI-01-0145-FEDER-031826, PREDICT PTDC/CCI-CIF/29877/2017 and internal IT project, QBigData  PEst-OE/EEI/LA0008/2013, funded by FCT. 
NP thanks FCT grant CEECIND/04594/2017.
Support from FCT (Portugal) through Grant UID/CTM/04540/2013 is also acknowledged.
}

\bibliographystyle{unsrt}
\bibliography{mybibv2}
\end{document}